\documentclass[prb,preprint]{revtex4-1} 

\usepackage{amsmath}  
\usepackage{amsfonts} 
\usepackage{graphicx} 

\begin{document}

\title{Bound orbits and gravitational theory}

\author{Naresh Dadhich}
\email{nkd@iucaa.ernet.in} 
\altaffiliation{Centre for Theoretical Physics, Jamia Millia Islamia,
New Delhi 110025, India}

\altaffiliation{Inter-University Centre for
Astronomy \& Astrophysics, Post Bag 4, Pune 411 007, India}

\author{Sushant G. Ghosh}
\email{sgghosh2@jmi.ac.in}
\affiliation{Centre for Theoretical Physics, Jamia Millia Islamia,
New Delhi 110025, India}

\author{Sanjay Jhingan}
\email{sanjay.jhingan@gmail.com}
\affiliation{Centre for Theoretical Physics, Jamia Millia Islamia,
New Delhi 110025, India}


\date{\today}

\begin{abstract}
It can be easily shown that bound orbits around a static source can exist only in $4$ dimension and in none else for any long range force. This is so not only for Maxwell's electromagnetic and Newton's gravity but also for Einstein's gravitation theory. In contrast to Maxwell's electrodynamics and Newton's gravity, GR has a natural higher dimensional generalization in Lovelock gravity which remarkably admits bound orbits around a static black hole in all even $d=2N+2$ dimensions where $N$ is degree of Lovelock polynomial action. This is as general a result as Bertrand's theorem of classical mechanics in which existence of closed  orbits uniquely singles out the inverse square law for a long range central force. 
\end{abstract}

\maketitle 

For existence of bound orbits what is required is that repulsive centrifugal force should be able to counterbalance attractive central force at two radii to give turning points to orbit, may what source be (electric or gravity). That means effective potential, which is sum of repulsive and attractive parts, should have a minimum. That also indicates existence of stable circular orbit. That is existence of bound orbits is equivalent to existence of stable circular orbit. We know that centrifugal potential always goes as $1/r^2$ while attractive force obeying the Gauss law of flux conservation would go as $1/r^{n+1}$ where $n=d-2$ and $d$ is dimension of space. It then readily follows that the condition for minimum of effective potential requires $n<2$. Bound orbits could thus occur  only in $3$ space dimension and in none else. This is true for any long range force including both electric and gravitational force. In higher space dimension $>3$ , there could occur no bound orbits implying total absence of structures like atoms and planetary systems. That means no structures at all in the Universe and thereby no life either. Is that why the Universe we live in is $4$ dimensional \cite{4-d}? \\ 

The conditions for effective potential $V(r)$ to be minimum are $V(r)^{\prime}=0, V(r)^{\prime\prime} > 0$ where a prime denotes derivative relative to $r$. For a central attractive force, effective potential is given by 
 \begin{equation}
 V = - \frac{M}{r^n} + \frac{l^2}{r^2}. 
\end{equation}
It would be minimum whenever 
 \begin{equation}
 n(2-n) > 0, 
\end{equation}
 which clearly demands $n = d-2 <2$. That is space dimension cannot be anything other than $3$. This would be true for any force that obeys the Gauss law. Thus bound orbits can exist only in $3$ dimensional space. 
 
Unlike Maxwell's electrodynamics, Newtonian gravity has a relativistic generalization in Einstein's general relativity (GR).  It turns out that the question of existence of bound orbits is completely neutral to Newton and Einstein;i.e. no bound orbits in GR either for $d>4$, now $d$ is dimension of spacetime. This could be seen as follows.  We write  the effective potential as  
 \begin{equation}
 V^2 = f(r)\left(\frac{l^2}{r^2} +1\right), 
\end{equation}
for a radially symmetric static  metric,   
\begin{equation}
 ds^2 = -f(r) dt^2 + \frac{1}{f(r)}dr^2 + r^2 d\Omega_{d-2}^2. 
\end{equation}
This form of the metric is dictated by the requirement that radially falling photon experiences no acceleration - velocity of light remains constant in vacuum \cite{null}. Now the conditions for minimum of effective potential give 

\begin{equation}
 \frac{l^2}{r^2} = \frac{rf^{\prime}}{2f - rf^{\prime}}
\end{equation}
and 
\begin{equation}
 rff^{\prime\prime} - 2 r{f^{\prime}}^2 + 3 f f^{\prime} >0.  \label{condition}
\end{equation}
For Einstein solution for a $d$ dimensional static black hole \cite{tengi}, we have 
\begin{equation}
  f(r) = 1 - M/r^n,
\end{equation}
where $n=d-3$. Then we have 
\begin{equation}
 \frac{l^2}{r^2} = \frac{nM}{2r^n-(n+2)M}
\end{equation}
which gives the existence threshold for circular orbit 
\begin{equation}
 r_{ex} > r_{ph} = \left(\frac{n+2}{2}M\right)^{1/n}
 \end{equation} 
where $r_{ph}$ is radius of photon circular orbit. This marks the existence threshold as no circular orbit can exist inside photon circular orbit. The stability threshold is given by 
 \begin{equation}
 r_{st} = \left(\frac{n+2}{2-n}M\right)^{1/n}.
\end{equation}
Clearly  $n  = d-3 < 2$. That is spacetime dimension cannot be any other than $4$. Thus for GR too bound orbits can exist only in $4$ dimension and in none else. So far as exietence of bound orbits is concerned both Newtonian gravity and GR are on the same footing. However in the latter orbits have a minimum radius for  inner turning point which is defined by the radius of photon circular orbit.  \\

Unlike Maxwell's electrodynamics, GR has the most natural generalization in higher dimensions, known as Lovelock gravity which includes GR in the linear order. Its action is a homogeneous polynomial in Riemann curvature yet it has the remarkable unique property that the equation of motion for field remains second order. If the polynomial degree is $N$, $N=1$ corresponds to GR and $N=2$ to Gauss-Bonnet, and so on. For a given $N$, $d \geq 2N+1$ and hence it is truly a higher dimensional generalization. In GR, gravity is kinamatic in $3$ and it becomes dynamic in $4$ dimension. Could this be a general gravitational property in higher dimensions;i.e. kinematic in odd and dynamic in even dimensions? This is precisely what has recently been established for pure Lovelock gravity \cite{dgj}. By pure we mean the action involves a single $N$th order polynomial with no summation over $N$ as for Einstein-Lovelock and in particular it is free of Einstein-Hilbert linear term $R$. Further it is possible to define an $N$th order analogue of Riemann curvature with the property that trace of its Bianchi derivative vanishes identically and that yields a second rank symmetric divergence free tensor which is the same as the one that comes from variation of the corresponding $N$th order Lovelock action \cite{bianchi}. Now gravity is kinematic in all odd $d=2N+1$ dimensions because $R^{(N)}_{ab}=0$ implies $R^{(N)}_{abcd}=0$. That is Lovelock vacuum is Lovelock flat in $2N+1$ dimension, however it won't be Riemann flat \cite{dgj}. Thus pure Lovelock gravity is always kinematic in odd $d=2N+1$ and becomes dynamic in even $d=2N+2$ dimension. This is a universal gravitational feature which has been established for the first time in Ref \cite{dgj}. \\

It has been proposed and strongly articulated \cite{n1} that proper gravitational equation in higher dimensions is pure Lovelock equation, 
\begin{equation}
G^{(N)}_{ab} = -\Lambda g_{ab} + \kappa T_{ab} 
\end{equation}
where $G^{(N)}_{ab}$ is defined as \cite{bianchi}
\begin{equation}
G^{(N)}_{ab} = N(R^{(N)}_{ab} - 1/2 R^{(N)} g_{ab}). 
\end{equation}
Einstein-Gauss-Bonnet vacuum solution for a static black hole in $5$ dimension was obtained by Boulware and Deser \cite{bd} which was followed by general Einstein-Lovelock solution for any $N$ \cite{whitt}. The ultimate equation to be solved is an algebraic $N$th degree polynomial which cannot in general be solved for $N>4$. However for pure Lovelock, there is no such difficulty \cite{lov-sol} and we have the general solution for $d=2N+2$ and it is given by 
\begin{equation}
 f(r)= 1-\left(\frac{M}{r}\right)^{1/N} . 
\end{equation}
Here $n=1/N$ which would always satisfy the required condition $n<2$ for existence of bound orbits. Thus bound orbits would always exist as also shown in Fig. for all even $2N+2$ dimensions in pure Lovelock gravity. This question is however not pertinent for odd $2N+1$ dimension as gravity is kinematic there. We have set $\Lambda = 0$ as it has no relevance in the region wher bound orbits are being sought. This is is indeed very significant distinguishing feature of pure Lovelock gravity which is not shared by Einstein and even Einstein-Lovelock as well as by any other modifications of GR. \\

For pure Lovelock case, let us for transparency and clarity give the relevant expressions explicitly and they read as  
\begin{equation}
 V^2 = \left[1 - \left(\frac{M}{r}\right)^{1/N}\right] \left(\frac{l^2}{r^2}+1\right),
\end{equation}
and 
\begin{equation}
 \frac{l^2}{r^2} = \frac{M^{1/N}}{\left(2N r^{1/N} - (2N+1)M^{1/N}\right)}
\end{equation}
which gives the existence threshold 
\begin{equation}
r_{ex} >  r_{ph} = \left(\frac{2N+1}{2N} \right)^N M.
\end{equation}
For stability threshold we have  
\begin{equation}
(V^2)^{\prime \prime} = \frac{2M^{1/N}((2N-1)r^{1/N}-(2N+1)M^{1/N})}{N r^{(2N+1)/N}(2Nr^{1/N}-(2N+1)M^{1/N})}.
\end{equation}
and it is given by 
\begin{equation}
 r_{st} = \left(\frac{2N+1}{2N-1} \right)^N M.
\end{equation}

For $N=1$ we have the familiar limits, $r_{ph} = 3/2 M$ and $r_{st} = 3M$. \\

\begin{figure}
       \centering
               \includegraphics[width=0.4\textwidth]{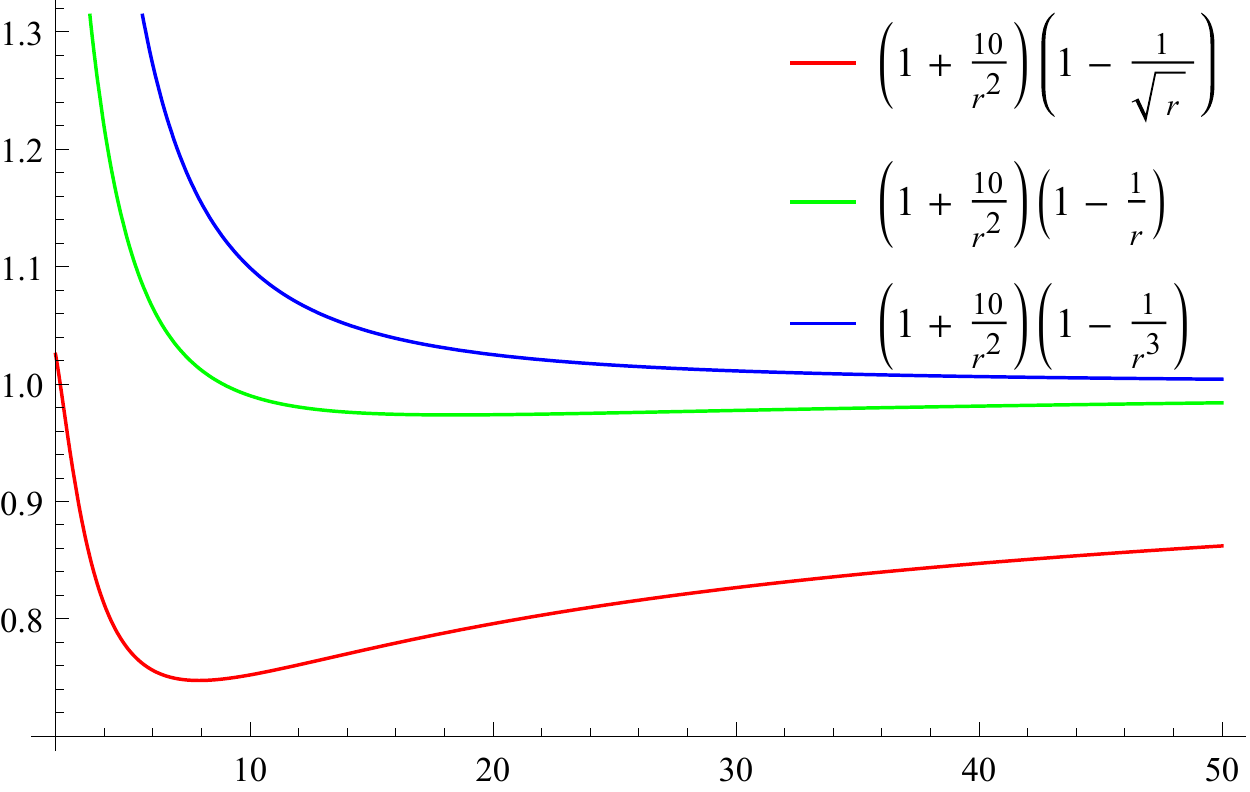}
       \caption{Shape of the potential in Einstein and Gauss-Bonnet theory.}
       \label{fig:fig}
\end{figure}

This is an interesting discriminator between Einstein and all its generalizations and pure Lovelock gravity. If bound orbits are to exist in all even dimensions, it can only be pure Lovelock gravity and none else. This is in fact as general a result as Bertrand's theorem of classical mechanics which by demand of existence of closed orbits uniquely singles out the inverse square law for a long range central force. We could thus state our result as a theorem as follows: \\

\emph{Theorem: If bound orbits are to exist in all even $2N+2$ dimensions, it can only be pure Lovelock gravity and none else where $N$ is degree of Lovelock action.}

\begin{acknowledgments}   SGG would like to thank
University Grant Commission (UGC) for major research project grant NO. F-39-459/2010(SR) and to IUCAA, Pune for kind hospitality while part of this work was being done.
SJ acknowledges support under UGC minor research project (42-1068/2013(SR).
\end{acknowledgments}



\begin{thebibliography}{99}

\bibitem{4-d} Dadhich N., Why do we live in four dimension? [arXiv:0902.0205].

\bibitem{null} Dadhich N., Einstein is Newton with space curved, [arXiv:12060635].

\bibitem{tengi} Tengherlini F. R., Nuovo Cimento {\bf27}, 636 (1963).

\bibitem{dgj} Dadhich N., Ghosh S. G. and Jhingan S., Phys.\ Lett.\ B {\bf 711} 196, (2012),
  [arXiv:1202.4575 [gr-qc]]

\bibitem{bianchi} Dadhich N., Pramana {\bf74}, 875 (2010), [arXiv:0802.3034].

\bibitem{n1} Dadhich N., Gravitational equation in higher dimensions, Proceedings of Relativity and Gravitation: 100 years after Einstein in Prague, June 25-28 (2012), [arXiv:1210.3022].

\bibitem{bd} Boulware D. G. and Deser S., Phys. Rev. Lett. {\bf 55}, 2656 (1985).

\bibitem{whitt} Wheeler J. T., Nucl.Phys. {\bf B268}, 737 (1986); {\bf B273}, 732 (1986); B. Whitt, Phys. Rev. {\bf D38}, 3000 (1988). 

\bibitem{lov-sol} Cai R. and Ohta N., Phys. Rev. D {\bf 74}, 064001 (2006) [hep-th/0604088]; Cai R., L-Ming Cao, Hu Y. and Kim S., Phys. Rev. {\bf 78}, 124012 (2008), [arXiv:0810.2610]; Dadhich N., Math Today {\bf 26}, 37 (2011) [arXiv:1006.0337]; Dadhich N., Pons J. M. and Prabhu K., Gen. Rel. Grav. {\bf 45}, 1131 (2013),
  [arXiv:1201.4994 [gr-qc]].



\end{thebibliography}
\end{document}